\documentclass[12pt]{article}
\usepackage{amssymb,amsmath}
\usepackage{epsfig}
\usepackage{graphicx}

\begin{document}
\title{Hubbard Hamiltonian in the dimer representation. Large $U$ limit}
\author{M.Matlak\thanks{e-mail:matlak@us.edu.pl}, J.Aksamit, B.Grabiec\\
Institute of Physics, Silesian University,\\
4 Uniwersytecka, PL-40-007 Katowice,Poland\\
and\\
W.Nolting\\
Institute of Physics, Humboldt University,\\
110 Invalidenstr., D-10115 Berlin, Germany}
\maketitle
\begin{abstract}
We formulate the Hubbard model for the simple cubic lattice in the
representation of interacting dimers applying the exact solution of the
dimer problem. By eliminating from the considerations unoccupied dimer
energy levels in the large $U$ limit (it is the only assumption) we
analytically derive the Hubbard Hamiltonian for the dimer (analogous to the
well-known $t-J$ model), as well as, the Hubbard Hamiltonian for the crystal
as a whole by means of the projection technique. Using this approach we can
better visualize the complexity of the model, so deeply hidden in its
original form. The resulting Hamiltonian is a mixture of many multiple
ferromagnetic, antiferromagnetic and more exotic interactions competing one
with another. The interplay between different competitive interactions has a
decisive influence on the resulting thermodynamic properties of the model,
depending on temperature, model parameters and assumed average number of
electrons per lattice site. A simplified form of the derived Hamiltonian can
be obtained using additionally Taylor expansion with respect to $x=\frac{t}{U}$ ($t$-hopping integral between nearest neighbours, $U$-Coulomb repulsion).
As an example, we present the expansion including all terms proportional to $t$ and to $\frac{t^2}U$ and we reproduce the exact form of the Hubbard
Hamiltonian in the limit $U\rightarrow \infty $.

The nonperturbative approach, presented in this paper, can, in principle, be
applied to clusters of any size, as well as, to another types of model
Hamiltonians.
\end{abstract}

\section{Introduction}

The single-band Hubbard model, Ref. [1], plays in the solid state physics a
similar principal role as the hydrogen atom in the atomic physics. This
explains a continuous interest in its properties. The Hubbard Hamiltonian
reads
\begin{equation}
H=\sum\limits_{i,j,\sigma }t_{i,j}c_{i,\sigma }^{+}c_{j,\sigma
}+U\sum\limits_in_{i,\uparrow }n_{i,\downarrow }.
\end{equation}
Here $c_{i,\sigma }$ $(c_{i,\sigma }^{+})$ are annihilation (creation)
operators of an electron with spin $\sigma =\uparrow $, $\downarrow $ in the
Wannier representation at the lattice site $\mathbf{R}_i$ and $n_{i,\sigma
}=c_{i,\sigma }^{+}c_{i,\sigma }$. Moreover, $t_{i,j}$ is the hopping
integral between different lattice sites $i$ and $j$ ($t_{i,i}=0)$ and $U$
is the intrasite Coulomb repulsion. The Bloch conduction band energy $%
\varepsilon _{\mathbf{k}}$ is given by

\begin{equation}
\varepsilon _{\mathbf{k}}=\sum\limits_{i-j}t_{i,j}e^{-i\mathbf{k\cdot (R}_i-%
\mathbf{R}_j)}.
\end{equation}
In the following we restrict ourselves to the simple cubic (sc) lattice and
assume that

\[
t_{i,j}=\left\{ 
\begin{array}{ll}
-t & i,j\mbox{-nearest neighbours} \\ 
0 & \mbox{otherwise}.
\end{array}
\right. 
\]
Then

\begin{equation}
\varepsilon _{\mathbf{k}}=-2t(\cos k_xa+\cos k_ya+\cos k_za).
\end{equation}

The hopping parameter $t$ is simply related to the bandwidth $W$ of the
Bloch band (2), e. g. $W=12t$ for the sc lattice$.$ The interplay between
the two model parameters, $W$ and $U,$ is decisive for the properties of the
model resulting in strong electron correlations, leading to band magnetism
(see e.g. Refs [2], [3] for a review), insulator-to-metal transition (see
e.g. Refs [3], [4] and papers cited therein) and high-$T_c$
superconductivity (negative $U$-model, see e.g. Ref. [5]). The Hubbard model
very often plays also a role of a submodel for many other more complicated
models (as e.g. Anderson model (see Ref. [6]), s-f model (see Refs [7,8])
and so on). Especially interesting, but difficult to handle are the
properties of the Hubbard model in such a regime of the model parameters
where the bandwidth $W$ is comparable to the Coulomb repulsion $U$. In a
large number of papers [9-30] the authors tried to solve this model using
many sophisticated methods. The exact solution, however, does not exist till
now. Many authors tried to change the situation in this field by introducing
the expansion parameter $x=\frac{t}{U}$ $(x\ll 1)$. This idea (cf Refs
[31,32]) consists in replacing ''difficult physics'' connected with the
model by ''difficult mathematics'' obtained by a laborious expansion with
respect to $x.$ Different methods connected with this problem have been
applied as e.g. the perturbation expansion (see Refs [18], [21]), canonical
transformation (see Refs [9], [12], [17], [22], [23]) or ab initio
derivations (see Refs [25], [28-30]). Most of the methods, leading to the $%
t-J$ model (or generalized $t-J$ model) are also summarized in Refs [27],
[4].

The goal of the present paper is just to show that we can take another,
nonperturbative way. In the first step we divide the crystal lattice into a
set of interacting dimers. In other words, we can rewrite the Hubbard
Hamiltonian (1) for the sc lattice (see Fig. 1) in the equivalent form 
\footnote{%
The dimer Fourier transformation
\par
$c_{I,1,\sigma }=\frac 1{\sqrt{N}}\sum\limits_{\mathbf{k}}c_{\mathbf{k,}%
\sigma }e^{i\mathbf{k\cdot R}_{I,1}},c_{I,2,\sigma }=\frac 1{\sqrt{N}%
}\sum\limits_{\mathbf{k}}c_{\mathbf{k,}\sigma }e^{i\mathbf{k\cdot R}_{I,2}}$%
\par
where $N$ is the number of lattice points, applied to (4), gives the well
known result
\par
$H=\sum\limits_{\mathbf{k,\sigma }}\varepsilon _{\mathbf{k}}n_{\mathbf{k,}%
\sigma }+\frac UN\sum\limits_{\mathbf{k,k}^{\prime },\mathbf{q}}c_{\mathbf{%
k+q,}\uparrow }^{+}c_{\mathbf{k,}\uparrow }c_{\mathbf{k}^{\prime }-\mathbf{q,%
}\downarrow }^{+}c_{\mathbf{k}^{\prime }\mathbf{,}\downarrow }$ with $%
\varepsilon _{\mathbf{k}}$ given by (3).}:

\begin{equation}
\begin{array}{ll}
H= & \sum\limits_IH_I^d-t\sum\limits_{I,\sigma }(c_{I,2,\sigma
}^{+}c_{I+1,1,\sigma }+c_{I+1,1,\sigma }^{+}c_{I,2,\sigma }) \\
& -t\sum\limits_{I\neq J,\sigma }(c_{I,1,\sigma }^{+}c_{J,1,\sigma
}+c_{I,2,\sigma }^{+}c_{J,2,\sigma })
\end{array}
\end{equation}
where

\begin{equation}
H_I^d=-t\sum\limits_{\sigma} (c_{I,1,\sigma }^{+}c_{I,2,\sigma
}+c_{I,2,\sigma }^{+}c_{I,1,\sigma })+U(n_{I,1,\uparrow }n_{I,1,\downarrow
}+n_{I,2,\uparrow }n_{I,2,\downarrow }).
\end{equation}
The indices $I$ and $J$ enumerate the dimers and $H_I^d$ is the dimer
Hamiltonian. The second term in (4) describes the hopping between nearest
dimers in the $z$-direction, the third one represents the hopping between
nearest dimers (($y,z$)-plane) and between different dimer planes (see Fig.
1).
\begin{figure}[h]
\noindent
\includegraphics[angle=-90,scale=0.45]{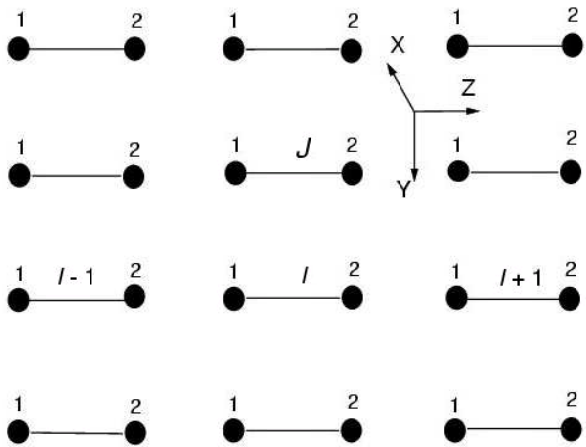}
\caption{\footnotesize {A plane of the Hubbard
dimers. The crystal is considered as a collection of parallel lying dimer
planes.}}
\end{figure}


The equivalent form of the Hubbard Hamiltonian (4) is especially suitable
because we can apply in the following the exact solution of the dimer
problem (5). In the next step we express the construction operators $%
a_{\sigma}$ ($a_{\sigma}^{+}$) as linear combinations of the transition
operators between different dimer states. This, in turn, allows to find the
exact dimer representation of the whole Hamiltonian (4). The space of the
dimer eigenvectors consists, however, of two subspaces. One of them
corresponds to the lowest lying energy levels, the second one contains the
levels with energies which in the large $U$ limit take on large, positive
values. These levels in a reasonable temperature range cannot be occupied by
electrons and therefore we can exclude them from further considerations. It
is interesting to note that this approch, without any additional
assumptions, applied to the dimer Hamiltonian (5) produces the analogy of
the well known $t-J$ model (cf Refs [31], [32]). A similar approach can be
applied to the Hamiltonian (4), describing the crystal as a whole. By
eliminating from the considerations the unoccupied dimer energy levels in
the large $U$ limit (it is the only assumption) with the use of the
projection technique we can find in a straightforward way the final form of
the Hamiltonian in this limit without using perturbation expansion or
canonical transformation. The resulting Hamiltonian, obtained in this way,
is very complicated. It, however, explicitly shows all possible magnetic,
nonmagnetic or more complex competitive interaction processes very deeply
hidden in the original Hamiltonian written in the site representation (1).
It is the aim of this paper just to reveal these important but normally
invisible elementary interactions. One important advantage might be that a
given approach to the unsolvable Hubbard problem can be tested with respect
to the types of neglected interaction processes. Besides, the new and
straightforward method, presented in this paper, can easily be adopted to
clusters of any size and also to another types of model Hamiltonians.

The paper is organized as follows. In Sec. 2 we find the exact solution of
the Hubbard dimer (5) and give the exact expressions for the annihilation
operators $c_{I,1(2),\sigma }$ in the dimer representation. In Sec. 3 we
derive the Hubbard dimer Hamilonian (5) in the large $U$ limit. With the use
of the projection technique onto the lowest lying dimer states we derive in
Sec. 4 the Hubbard Hamiltonian for the crystal in this limit (central
formula of this paper). A simplified version of the derived Hamiltonian with
the use of the Taylor expansion with respect to $x=\frac{t}{U}$ $(x\ll 1)$
and the case $U\rightarrow \infty $ is discussed in Sec. 5.

\section{Exact solution of the Hubbard dimer}

The eigenvalues and eigenvectors of the dimer Hamiltonian (5) can be found
by using a standard procedure (cf Refs [33-35]). We start with the vectors $%
|n_{1,\uparrow },n_{1,\downarrow };n_{2,\uparrow },n_{2,\downarrow }\rangle $
$(n_{i,\sigma }=0,1;$ $i=1,2$; $\sigma =\uparrow ,\downarrow )$ which
constitute the Fock basis of the single-dimer space of states:

\begin{equation}
\begin{array}{lll}
\begin{array}{l}
\begin{array}{l}
|0\rangle =|0,0;0,0\rangle ,
\end{array}
\\ 
\\ 
\begin{array}{l}
|11\rangle =|1,0;0,0\rangle , \\ 
|12\rangle =|0,1;0,0\rangle , \\ 
|13\rangle =|0,0;1,0\rangle , \\ 
|14\rangle =|0,0;0,1\rangle ,
\end{array}
\end{array}
& 
\begin{array}{l}
|21\rangle =|1,1;0,0\rangle , \\ 
|22\rangle =|1,0;1,0\rangle , \\ 
|23\rangle =|1,0;0,1\rangle , \\ 
|24\rangle =|0,1;1,0\rangle , \\ 
|25\rangle =|0,1;0,1\rangle , \\ 
|26\rangle =|0,0;1,1\rangle ,
\end{array}
& 
\begin{array}{l}
\begin{array}{l}
|31\rangle =|0,1;1,1\rangle , \\ 
|32\rangle =|1,0;1,1\rangle , \\ 
|33\rangle =|1,1;0,1\rangle , \\ 
|34\rangle =|1,1;1,0\rangle ,
\end{array}
\\
\\ 
\begin{array}{l}
|4\rangle =|1,1;1,1\rangle .
\end{array}
\end{array}
\end{array}
\end{equation}
Starting with the vectors (6) we easily get the eigenvalues $E_\alpha $ and
the eigenvectors $|E_{\alpha} \rangle $ of the Hubbard dimer (5). Here, we
only mention that the space of 16 eigenvectors $|E_{\alpha} \rangle $ can be
devided into some subspaces numbered by $n=\sum\limits_{i,\sigma
}n_{i,\sigma }$. The subspace, belonging to $n=0$ and $n=4$ is 1-dimensional
($E_0=0,$ $|E_0\rangle =|0\rangle ;$ $E_4=2U,$ $|E_4\rangle =|4\rangle )$.
There are, however, two 2-dimensional subspaces corresponding to $n=1$ (as
e.g. $E_{11}=-t,$ $|E_{11}\rangle =\frac 1{\sqrt{2}}(|11\rangle +|13\rangle
),$ etc.) and two 2-dimensional subspaces corresponding to $n=3$ (as e.g. $%
E_{31}=t+U,$ $|E_{31}\rangle =\frac 1{\sqrt{2}}(|31\rangle +|33\rangle ),$
etc.). The subspace, belonging to $n=2$ consists of one 4-dimensional
subspace (as e.g. $E_{21}=0,$ $|E_{21}\rangle =\frac 1{\sqrt{2}}(|23\rangle
+|24\rangle ),$ etc.) and two 1-dimensional subspaces ($E_{25}=E_{26}=0,$ $%
|E_{25}\rangle =|22\rangle ,$ $|E_{26}\rangle =|25\rangle $). The complete
set of the eigenvalues $E_{\alpha} $ and eigenvectors $|E_{\alpha} \rangle $
of the Hubbard dimer is given in the Appendix A. It allows to express the
dimer Hamiltonian (5) in the equivalent form

\begin{equation}
H^d=\sum\limits_{\alpha} E_{\alpha} |E_{\alpha} \rangle \langle E_{\alpha} |.
\end{equation}

The next important step in our calculations is the possibility to express
the annihilation operators $c_{1(2),\sigma }$ as linear combinations of
transition operators between dimer states $P_{\alpha ,\beta }=|E_{\alpha}
\rangle \langle E_{\beta} |.$ This procedure can easily be performed when
acting with the annihilation operators on the basis vectors (6) and using
the reciprocal relations to (A.1). In this way we can obtain the dimer
representation of the annihilation (creation) operators, given in Appedix B.
After this operation we can insert so prepared $c_{I,1(2),\sigma }$ $%
(c_{I,1(2),\sigma }^{+})$ into the Hubbard Hamiltonian (4) to obtain the
Hubbard model in the dimer representation for the sc lattice. This
representation will be used later to derive the Hubbard Hamiltonian in the
large $U$ limit (see Sec. 4).

\section{Hubbard dimer for large $U$}

Looking at the eigenvalues (A.1) of the Hubbard dimer it is easy to see that
for large $U$ $(U\gg t)$ the energies $E_{\alpha \text{ }%
}=E_{22},E_{23},E_{31},E_{32},E_{33},E_{34}$ and $E_4$ take on large,
positive values, producing in the partition function the terms which can
practically be neglected. It means that the mentioned energies are not
occupied in the reasonable temperature range ($1$ $eV\sim 11604.5$ $K$) and
can be excluded from our considerations. Therefore the dimer Hamiltonian,
given by (7), reduces to

\begin{equation}
\overline{H}^d=-tP_{11,11}+tP_{12,12}-tP_{13,13}+tP_{14,14}+\left( -C+\frac
U2\right) P_{24,24}.
\end{equation}
To bring this expression into a compact (second quanization) form we
introduce the Hubbard operators

\begin{eqnarray}
a_{i,\sigma } &=&c_{i,\sigma }(1-n_{i,-\sigma }), \\
b_{i,\sigma } &=&c_{i,\sigma }n_{i,-\sigma }
\end{eqnarray}
and spin operators

\begin{eqnarray}
S_i^z &=&\frac 12(n_{i,\uparrow }-n_{i,\downarrow })=\frac 12(n_{i,\uparrow
}^a-n_{i,\downarrow }^a), \\
S_i^{+} &=&c_{i,\uparrow }^{+}c_{i,\downarrow }=a_{i,\uparrow
}^{+}a_{i,\downarrow }, \\
S_i^{-} &=&c_{i,\downarrow }^{+}c_{i,\uparrow }=a_{i,\downarrow
}^{+}a_{i,\uparrow }
\end{eqnarray}
where $n_{i,\sigma }^a=a_{i,\sigma }^{+}a_{i,\sigma }$ $(i=1,2;$ $\sigma
=\uparrow $, $\downarrow ).$

With the use of (9)-(13) the Hamiltonian of the Hubbard dimer (8) for large $%
U$ can be presented in the form (we introduce the omitted earlier dimer
index $I$)

\begin{eqnarray}
\overline{H}_I^d &=&-t\sum\limits_{\sigma} [a_{I,1,\sigma }^{+}a_{I,2,\sigma
}+a_{I,2,\sigma }^{+}a_{I,1,\sigma }]  \nonumber \\
&&  \nonumber \\
&&+\frac{4t^2}{U\sqrt{1+(\frac{4t}U)^2}}[\overrightarrow{S}_{I,1}\cdot 
\overrightarrow{S}_{I,2}-\frac{n_{I,1}^an_{I,2}^a}4]  \nonumber \\
&&  \nonumber \\
&&+\frac{t^2(1-\sqrt{1+(\frac{4t}U)^2})}{U(1+\sqrt{1+(\frac{4t}U)^2})\sqrt{%
1+(\frac{4t}U)^2}}[2(b_{I,1,\uparrow }^{+}a_{I,1,\downarrow
}^{+}a_{I,2,\downarrow }b_{I,2,\uparrow }  \nonumber \\
&&  \nonumber \\
&&+b_{I,2,\uparrow }^{+}a_{I,2,\downarrow }^{+}a_{I,1,\downarrow
}b_{I,1,\uparrow }) \\
&&  \nonumber \\
&&+(1-n_{I,1}^a)n_{I,2}^b+(1-n_{I,2}^a)n_{I,1}^b-n_{I,1}^bn_{I,2}^b] 
\nonumber \\
&&  \nonumber \\
&&+\frac{t(1-\sqrt{1+(\frac{4t}U})^2)}{2\sqrt{1+(\frac{4t}U)^2}}%
\sum\limits_{\sigma} \sum\limits_{\alpha =1,2}[a_{I,\alpha ,\sigma }^{+}b_{I,%
\overline{\alpha },\sigma }+b_{I,\alpha ,\sigma }^{+}a_{I,\overline{\alpha }%
,\sigma }]  \nonumber
\end{eqnarray}

where $n_{I,\alpha }^{a,b}=\sum\limits_{\sigma} n_{I,\alpha ,\sigma }^{a,b},$
$n_{I,\alpha ,\sigma }^b=b_{I,\alpha ,\sigma }^{+}b_{I,\alpha ,\sigma
}=n_{I,\alpha ,\sigma }n_{I,\alpha ,-\sigma }$ $(\alpha =1,2)$, $\overline{%
\alpha }=1$ when $\alpha =2$ and $\overline{\alpha }=2$ when $\alpha =1$. It
is very important to stress that the formula (14) has been obtained in a
nonperturbative way, starting from the exact form of the dimer Hamiltonian
(see (5) or (7)), and excluding from the considerations unoccupied dimer
energy levels in the large $U$ limit. Let us note that using the Taylor
expansion in (14) with respect to $x=\frac{t}{U}$ and retaining the terms
proportional to $\frac{t^2}U$ we obtain

\begin{equation}
\overline{H}_I^d=-t\sum\limits_\sigma [a_{I,1,\sigma }^{+}a_{I,2,\sigma
}+a_{I,2,\sigma }^{+}a_{I,1,\sigma }]+\frac{4t^2}U[\overrightarrow{S}%
_{I,1}\cdot \overrightarrow{S}_{I,2}-\frac{n_{I,1}^an_{I,2}^a}4].
\end{equation}
The formula (15) is the well-known $t-J$ model for the Hubbard dimer where
the first part in (15), similarly to (14), represents the exact form of the
dimer Hamiltonian (5) in the limit $U\rightarrow \infty $.

The same result (14) can also be obtained applying a more general approach
which will be used later to derive the Hubbard Hamiltonian for large $U$ in
the case of a crystal. Let us note that after elimination of the unoccupied
levels the subspace of the eigenvectors of the Hubbard dimer (A.1) for large 
$U$ consists of the following eigenvectors: $|E_0\rangle $, $|E_{11}\rangle $%
, $|E_{12}\rangle $, $|E_{13}\rangle $, $|E_{14}\rangle $, $|E_{21}\rangle $%
, $|E_{24}\rangle $, $|E_{25}\rangle $, and $|E_{26}\rangle $. It means that
we can define a projection operator onto this subspace

\begin{equation}
\begin{array}{ll}
P_I= & 
P_{0,0}^{(I)}+P_{11,11}^{(I)}+P_{12,12}^{(I)}+P_{13,13}^{(I)}+P_{14,14}^{(I)}
\\ 
& +P_{21,21}^{(I)}+P_{24,24}^{(I)}+P_{25,25}^{(I)}+P_{26,26}^{(I)}
\end{array}
\end{equation}
which, in the second quantization form, reads

\begin{equation}
\begin{array}{ll}
P_I= & 1-\frac 12(n_{I,1}^b+n_{I,2}^b)+\frac 14n_{I,1}^bn_{I,2}^b \\ 
&  \\ 
& 
\begin{array}{ll}
& +\frac 18(1-\frac 1{\sqrt{1+(\frac{4t}U)^2}})[4(\overrightarrow{S}%
_{I,1}\cdot \overrightarrow{S}_{I,2}-\frac 14n_{I,1}^an_{I,2}^a) \\ 
&  \\ 
& +2(b_{I,1,\uparrow }^{+}a_{I,1,\downarrow }^{+}a_{I,2,\downarrow
}b_{I,2,\uparrow }+b_{I,2,\uparrow }^{+}a_{I,2,\downarrow
}^{+}a_{I,1,\downarrow }b_{I,1,\uparrow }) \\ 
&  \\ 
& +(1-n_{I,1}^a)n_{I,2}^b+(1-n_{I,2}^a)n_{I,1}^b-n_{I,1}^bn_{I,2}^b]
\end{array}
\\ 
&  \\ 
& +\frac t{U\sqrt{1+(\frac{4t}U)^2}}\sum\limits_{\sigma} \sum\limits_{\alpha
=1,2}(a_{I,\alpha ,\sigma }^{+}b_{I,\overline{\alpha },\sigma }+b_{I,\alpha
,\sigma }^{+}a_{I,\overline{\alpha },\sigma }).
\end{array}
\end{equation}
Now, the Hamiltonian (14) can also be obtained from (5) (or (7)) with the
use of the projection operator (16) or (17) using the relation

\begin{equation}
\overline{H}_I^d=P_IH_I^dP_I
\end{equation}
and applying a straightforward but laborious algebraic calculation.

\section{Hubbard model for large $U$}

The Hamiltonian of the whole crystal (4) can be expressed (similar to (7))
in the form 
\begin{equation}
H=\sum\limits_{\gamma} \overline{E}_{\gamma} |\overline{E}_{\gamma} \rangle
\langle \overline{E}_{\gamma} |
\end{equation}
with unknown energies $\overline{E}_{\gamma} $ and eigenvectors $|\overline{E%
}_{\gamma} \rangle $. We can, however, expand the eigenvectors $|\overline{E}%
_{\gamma} \rangle $ in the series of the dimer eigenvectors (see (A.1))

\begin{equation}
|\overline{E}_{\gamma} \rangle =\sum\limits_{\gamma _1,..,\gamma
_M}c_{\gamma _{1,}.._{,}\gamma _M}^{\gamma} |E_{\gamma _1}\rangle
..|E_{\gamma _M}\rangle
\end{equation}
assuming that the crystal consists of $M$ dimers. Using (19) and (20) we
obtain

\begin{equation}
H=\sum\limits_{\gamma} \overline{E}_{\gamma} \sum\limits_{\gamma
_1,..,\gamma _M}\sum\limits_{\gamma _1^{^{,}},..,\gamma _M^{,}}c_{\gamma
_{1,}.._{,}\gamma _M}^{\gamma} c_{\gamma _{1,}^{,}.._{,}\gamma
_M^{,}}^{\gamma *}|E_{\gamma _1}\rangle ..|E_{\gamma _M}\rangle \langle
E_{\gamma _1^{,}}|..\langle E_{\gamma _M^{,}}|.
\end{equation}

It is clear that to obtain the Hubbard Hamiltonian for large $U$ we have to
project (21) onto the subspace of the lowest lying dimer states with the use
of the projection operator

\begin{equation}
P=P_1P_2...P_M
\end{equation}
where $P_{I}$ is given by (16) or (17). In analogy to (14) we denote the
Hubbard Hamiltonian in the large $U$ limit by $\overline{H}$. Similar to
(18) we write

\begin{equation}
\overline{H}=PHP
\end{equation}

and instead of the form (21) for the Hamiltonian $H$ we can use (4). Taking
into account that $P^2=P$ $(P_I^2=P_I,$ $\left[ P_I,P_J\right] =0)$ we obtain

\begin{equation}
\begin{array}{ll}
\overline{H}= & P[\sum\limits_I\overline{H}_I^d-t\sum\limits_{I,\sigma
}\left( \overline{c}_{I,2,\sigma }^{+}\overline{c}_{I+1,1,\sigma }+\overline{%
c}_{I+1,1,\sigma }^{+}\overline{c}_{I,2,\sigma }\right) \\ 
& -t\sum\limits_{I\neq J,\sigma }\left( \overline{c}_{I,1,\sigma }^{+}%
\overline{c}_{J,1\sigma }+\overline{c}_{I,2,\sigma }^{+}\overline{c}%
_{J,2,\sigma }\right) ]\equiv P\overline{\overline{H}}
\end{array}
\end{equation}

where $\overline{H}_I^d$ is given by (14) and $(\alpha =1,2)$

\begin{equation}
\overline{c}_{I,\alpha ,\sigma }=c_{I,\alpha ,\sigma }P_I,
\end{equation}

\begin{equation}
\overline{c}_{I,\alpha ,\sigma }^{+}=P_Ic_{I,\alpha ,\sigma }^{+}.
\end{equation}

Applying the projection operator (16) or (17) to (B.1) - (B.4) and
introducing Hubbard- and spin operators (9) - (13) we obtain

\begin{equation}
\overline{c}_{I,1,\uparrow }=\underline{\underline{a}}_{I,1,\uparrow }+\beta
\left[ S_{I,2}^za_{I,1,\uparrow }+S_{I,2}^{-}a_{I,1,\downarrow }\right]
-\delta \left[ S_{I,1}^za_{I,2,\uparrow }+S_{I,1}^{-}a_{I,2,\downarrow
}\right] ,
\end{equation}

\begin{equation}
\overline{c}_{I,1,\downarrow }=\underline{\underline{a}}_{I,1,\downarrow
}-\beta \left[ S_{I,2}^za_{I,1,\downarrow }-S_{I,2}^{+}a_{I,1,\uparrow
}\right] +\delta \left[ S_{I,1}^za_{I,2,\downarrow
}-S_{I,1}^{+}a_{I,2,\uparrow }\right]
\end{equation}

where the corresponding expressions for $\overline{c}_{I,2,\sigma }$ ($%
\sigma =\uparrow ,\downarrow $) can easily be obtained by changing the
internal dimer index $1\Leftrightarrow 2$ in (27) and (28).

The new operators $\underline{\underline{a}}_{I,\alpha ,\sigma }\left(
\alpha =1,2;\sigma =\uparrow ,\downarrow \right) $ are introduced to obtain
a relatively compact form of (27) and (28). They are defined as follows

\begin{equation}
\begin{array}{ll}
\underline{\underline{a}}_{I,\alpha ,\sigma }= & \underline{a}_{I,\alpha
,\sigma }+\beta \left( \underline{b}_{I,\alpha ,\sigma }+a_{I,\alpha
,-\sigma }^{+}a_{I,\overline{\alpha },-\sigma }b_{I,\overline{\alpha }%
,\sigma }\right) \\ 
& +\delta \left[ \underline{b}_{I,\overline{\alpha },\sigma }+a_{I,\overline{%
\alpha },-\sigma }^{+}a_{I,\alpha ,-\sigma }b_{I,\alpha ,\sigma }+a_{I,%
\overline{\alpha },\sigma }\frac{n_{I,\alpha }^a}2\right]
\end{array}
\end{equation}

where

\begin{equation}
\underline{a}_{I,\alpha ,\sigma }=a_{I,\alpha ,\sigma }\left( 1-\frac \beta
2n_{I,\overline{\alpha }}^a-\frac 12n_{I,\overline{\alpha }}^b\right) ,
\end{equation}

\begin{equation}
\underline{b}_{I,\alpha ,\sigma }=b_{I,\alpha ,\sigma }\left( 1-n_{I,%
\overline{\alpha }}^a-\frac 12n_{I,\overline{\alpha }}^b\right)
\end{equation}

and

\begin{equation}
\beta =\frac 14\left( 1-\frac 1{\sqrt{1+\left( \frac{4t}U\right) ^2}}\right)
,
\end{equation}

\begin{equation}
\delta =\frac{t}{U}\frac 1{\sqrt{1+\left( \frac{4t}U\right) ^2}}.
\end{equation}

To write down the explicit form of the Hubbard Hamiltonian $\overline{%
\overline{H}}$ (see (24)) we have to insert (27) and (28) into $\overline{%
\overline{H}}$. This operation leads, however, to a very complicated form of 
$\overline{\overline{H}},$ given in the Appendix C (a simplified form of
this Hamiltonian is discussed in the next Section). Here again the only
assumption made to derive $\overline{\overline{H}}$ was the reduction of the
whole dimer space (A.1) to the subspace of the dimer eigenvectors $\left|
E_0\right\rangle $, $\left| E_{11}\right\rangle $, $\left|
E_{12}\right\rangle $, $\left| E_{13}\right\rangle $, $\left|
E_{14}\right\rangle $, $\left| E_{21}\right\rangle $, $\left|
E_{24}\right\rangle $, $\left| E_{25}\right\rangle $ and $\left|
E_{26}\right\rangle $, corresponding to the lowest lying dimer energy levels
because in the large $U$ limit only these levels can be occupied. The
Hamiltonian $\overline{\overline{H}}$, obtained in this way, contains many
competing, magnetic, nonmagnetic and more complex interactions. Among them
we can find a direct antiferromagnetic interaction generated by the term $%
\stackrel{\rightarrow }{S}_{I,1}\cdot \stackrel{\rightarrow }{S}_{I,2}$
(Heisenberg exchange interaction) multiplied by the positive coupling
constant. Such a term appears in $\overline{H}_I^d$ (see (C.6) and (14)).
Inside of $\overline{\overline{H}}$ (see (C.6) and (C.2) - (C.5)) a kind of
ferromagnetic interactions between spins from different dimers, represented
by (C.3), appears within the terms proportional to $\beta ^2$ and $\delta ^2$
(negative coupling constants). The antiferromagnetic interactions, however,
appear again in terms proportional to $\beta \delta $. There are also many
other magnetic, more exotic interactions, represented by (C.2), (C.4) and
(C.5), entering into (C.6). The situation is, however, much more complicated
when we consider the total Hamiltonian $\overline{H}$ (see (24)) in the
large $U$ limit. $\overline{H}$ differs from $\overline{\overline{H}}$ by
the multiplicative factor $P$ (a product of the projection operators $P_I$
(see (22) and (17)), standing on the left. Inside of each $P_I$ the
mentioned antiferromagnetic interaction also appears (see (17)). In other
words, the total Hamiltonian $\overline{H}$, we are interested in, is
actually a sum of the products of many competitive ferromagnetic,
antiferromagnetic and more exotic interactions. The thermodynamic properties
of the system, described by the Hamiltoniam $\overline{H}$ (24), are then a
result of the competition between all of them. Which interaction wins in
such a competition it certainly depends on temperature, model parameters ($%
t,U$) and on the average number of electrons per lattice site, determining
the chemical potential of the system.

The formalism presented in this paper is also applicable to a more
complicated decomposition of the Hubbard Hamiltonian (1) into a set od
interacting clusters consisting e.g. of one central atom and $z$ its nearest
neighbours. We, however, know (see e.g. Refs [36]-[39] and papers cited
therein) that, unfortunately, the mathematical problems in this case
exponentially grows up with the size of the cluster.

\section{Taylor expansion}

The complicated form of the Hubbard Hamiltonian $\overline{H}$ in the large $%
U$ limit (see (24), (C.6)) where $P$ is given by (22) (see also (17)) can
essentially be reduced when applying the Taylor expansion with respect to
the parameter $x=\frac{t}{U}\ll 1$. To do it we have to expand all the
coefficients in (14), (17) and (C.6) including also $\beta $ and $\delta $
(see (32), (33)). Such an expansion can be performed to any power of $x$,
however, the most simple form we obtain when we restrict ourselves to the
linear approximation, resulting in the terms proportional to $t$ and $\frac{%
t^2}U.$ The accuracy of this expansion can easily be verified assumming e.g.
a typical value of the ratio $\frac{W}{U}=\frac 15$ (or less). Because the
bandwidth of the conduction band for the sc lattice is $W=12t,$ it results
in a small value of the expansion parameter $x=\frac tU=\frac 1{60}$ in this
case. It, however, means that the linear approximation is quite reasonable
because all higher terms in the expansion, proportional to $x^n$ ($%
n=2,3,...),$ produce $60$ times smaller contribution.

To present the results of the Taylor expansion including all the terms
proportional to $t$ and $\frac{t^2}U$ let us first define several auxiliary
quantities

\begin{equation}
P^{(1)}=\prod\limits_{I=1}^MP_I^{(1)},
\end{equation}
\begin{equation}
P^{(2)}=\frac tU\sum\limits_{I=1}^MP_1^{(1)}\cdot ...\cdot
P_{I-1}^{(1)}P_I^{(2)}P_{I+1}^{(1)}\cdot ...\cdot P_M^{(1)}
\end{equation}
where

\begin{equation}
P_I^{(1)}=1-\frac 12\left( n_{I,1}^b+n_{I,2}^b\right) +\frac
14n_{I,1}^bn_{I,2}^b,
\end{equation}

\begin{equation}
P_I^{(2)}=\sum\limits_{\sigma} \sum\limits_{\alpha =1}^2\left( a_{I,\alpha
,\sigma }^{+}b_{I,\overline{\alpha },\sigma }+b_{I,\alpha ,\sigma }^{+}a_{I,%
\overline{\alpha },\sigma }\right)
\end{equation}
and ($\alpha =1,$ $2$)

\begin{equation}
\widetilde{a}_{I,\alpha ,\sigma }=a_{I,\alpha ,\sigma }\left( 1-\frac 12n_{I,%
\overline{\alpha }}^b\right) ,
\end{equation}

\begin{equation}
\widetilde{\widetilde{a}}_{I,\alpha ,\sigma }=\underline{b}_{I,\overline{%
\alpha },\sigma }+a_{I,\overline{\alpha },-\sigma }^{+}a_{I,\alpha ,-\sigma
}b_{I,\alpha ,\sigma }+a_{I,\overline{\alpha },\sigma }\frac{n_{I,\alpha }^a}%
2
\end{equation}
where $\underline{b}_{I,\alpha ,\sigma }$ is given by (31).

The Hamiltonian $\overline{\overline{H}}$ (C.6), including all the terms
proportional to $t$ and $\frac{t^2}U$, takes on the form

\begin{equation}
\overline{\overline{H}}=\overline{\overline{H}}^{(1)}+\overline{\overline{H}}%
^{(2)}
\end{equation}
where
\begin{eqnarray}
\overline{\overline{H}}^{(1)} &=&-t\sum\limits_{I,\sigma }\left[
a_{I,1,\sigma }^{+}a_{I,2,\sigma }+a_{I,2,\sigma }^{+}a_{I,1,\sigma }\right]
\nonumber \\[0.2cm]
&&-t\sum\limits_{I,\sigma }\left[ \widetilde{a}_{I,2,\sigma }^{+}\widetilde{%
a}_{I+1,1,\sigma }+\widetilde{a}_{I+1,1,\sigma }^{+}\widetilde{a}%
_{I,2,\sigma }\right] \\[0.2cm]
&&-t\sum\limits_{I\neq J,\sigma }\sum\limits_{\alpha =1}^2\widetilde{a}%
_{I,\alpha ,\sigma }^{+}\widetilde{a}_{J,\alpha ,\sigma }  \nonumber
\end{eqnarray}

and
\begin{eqnarray}
\overline{\overline{H}}^{(2)} &=&\frac{4t^2}U\sum\limits_I[\overrightarrow{S%
}_{I,1}\cdot \overrightarrow{S}_{I,2}-\frac 14n_{I,1}^an_{I,2}^a]  \nonumber
\\
&&-\frac{t^2}U\sum\limits_{I,\sigma }[\widetilde{a}_{I,2,\sigma }^{+}%
\widetilde{\widetilde{a}}_{I+1,1,\sigma }+\widetilde{\widetilde{a}}%
_{I,2,\sigma }^{+}\widetilde{a}_{I+1,1,\sigma }  \nonumber \\
&&+\widetilde{a}_{I+1,1,\sigma }^{+}\widetilde{\widetilde{a}}_{I,2,\sigma }+%
\widetilde{\widetilde{a}}_{I+1,1,\sigma }^{+}\widetilde{a}_{I,2,\sigma }]
\nonumber \\
&&-\frac{t^2}U\sum\limits_{I\neq J,\sigma }\sum\limits_{\alpha =1}^2[%
\widetilde{a}_{I,\alpha ,\sigma }^{+}\widetilde{\widetilde{a}}_{J,\alpha
,\sigma }+\widetilde{\widetilde{a}}_{I,\alpha ,\sigma }^{+}\widetilde{a}%
_{J,\alpha ,\sigma }] \\
&&+\frac{2t^2}U\sum\limits_I[\overrightarrow{S}_{I,2}\cdot (\overrightarrow{%
\underline{\underline{s}}}_{I,1;I+1,1}+\overrightarrow{\underline{s}}%
_{I+1,1;I,1})  \nonumber \\
&&+\overrightarrow{S}_{I+1,1}\cdot (\overrightarrow{\underline{\underline{s}}%
}_{I+1,2;I,2}+\overrightarrow{\underline{s}}_{I,2;I+1,2})]  \nonumber \\
&&+\frac{2t^2}U\sum\limits_{I\neq J}[\overrightarrow{S}_{I,1}\cdot (%
\overrightarrow{\underline{\underline{s}}}_{I,2;J,1}+\overrightarrow{%
\underline{s}}_{J,1;I,2})  \nonumber \\
&&+\overrightarrow{S}_{I,2}\cdot (\overrightarrow{\underline{\underline{s}}}%
_{I,1;J,2}+\overrightarrow{\underline{s}}_{J,2;I,1})].  \nonumber
\end{eqnarray}

The operators $\underline{\underline{s}}_{I,\mu ;J,\nu }^{z,\pm }$ and $%
\underline{s}_{I,\mu ;J,\nu }^{z,\pm }$ in (42) retain their forms
introduced in (C.1) but $\underline{\underline{a}}_{I,\mu ,\sigma }(%
\underline{\underline{a}}_{I,\mu ,\sigma }^{+})$ in (C.1) should actually be
replaced by $\widetilde{a}_{I,\mu ,\sigma }(\widetilde{a}_{I,\mu ,\sigma
}^{+})$, defined by (38). The total Hamiltonian $\overline{H}$ in the large $%
U$ limit (24) including all terms proportional to $t$ and $\frac{t^2}U$ can
thus be written in the form

\begin{equation}
\overline{H}=P^{(1)}(\overline{\overline{H}}^{(1)}+\overline{\overline{H}}%
^{(2)})+P^{(2)}\overline{\overline{H}}^{(1)}=P^{(1)}\overline{\overline{H}}%
^{(1)}+(P^{(1)}\overline{\overline{H}}^{(2)}+P^{(2)}\overline{\overline{H}}%
^{(1)}).
\end{equation}

The first part, $P^{(1)}\overline{\overline{H}}^{(1)}$, contains the terms
proportional to $t$ whereas the term in the parentheses is proportional to $%
\frac{t^2}U$ (see (35) and (42)). In the expression for $\overline{\overline{%
H}}^{(2)}$(see (42)) the first term describes the antiferromagnetic,
Heisenberg intradimer interaction (see also the second term in (15)). The
apearence of this term may suggest that such an interaction should also
arise between different dimers. This is of course the case. However, because
of applied procedure such terms do not appear explicitly. In the last
analysis our method treats all interactions within and between dimers, on
the same quality level, i.e. all interactions are taken into account. The
magnetic interdimer interactions, represented by the fourth and fifth term
in (42), have formally the same structure as the Heisenberg interactions but
instead of the scalar products of spin operators there are the products of
spin operators and ''hopping spin'' operators (defined in C.1). All the
terms presented in (42) are correct because they originate from the exact
decomposition of the Hubbard Hamiltonian (1) into a set of interacting
dimers (4) where each dimer problem has been exactly solved (exact dimer
representation of the construction operators after applying the projection
procedure, given by (27) and (28)). The total Hamiltonian $\overline{H}$
(43), we are interested in, is much more complicated than $\overline{%
\overline{H}}^{(1)}$ and $\overline{\overline{H}}^{(2)}$alone (see
(41)-(43)) because of the presence of the projection operators $P^{(1)}$ and 
$P^{(2)}$ (see (34)-(37)) in the expression (43).

It is also interesting to see what happens in the special case when taking
the limit $U\rightarrow \infty $. The second term in the parentheses of
Eq.(43) vanishes in this limit (cf (35) and (42)). Besides, each lattice
site cannot be at the same time occupied by two electrons what is equivalent
to the assumption ($\alpha =1,2)$

\begin{equation}
n_{I,\alpha ,\sigma }^b=b_{I,\alpha ,\sigma }^{+}b_{I,\alpha ,\sigma
}=n_{I,\alpha ,\sigma }n_{I,\alpha ,-\sigma }=0,
\end{equation}

\begin{equation}
n_{I,\alpha }^b=\sum\limits_{\sigma} n_{I,\alpha ,\sigma }^b=0
\end{equation}
and (cf (38))

\begin{equation}
\widetilde{a}_{I,\alpha ,\sigma }=a_{I,\alpha ,\sigma }.
\end{equation}
The total Hamiltonian (43) is thus given by a simple formula

\begin{equation}
\begin{array}{ll}
\overline{H}=P^{(1)}\overline{\overline{H}}= & -t\sum\limits_{I,\sigma
}\left[ a_{I,1,\sigma }^{+}a_{I,2,\sigma }+a_{I,2,\sigma }^{+}a_{I,1,\sigma
}\right] \\
&  \\
& -t\sum\limits_{I,\sigma }\left[ a_{I,2,\sigma }^{+}a_{I+1,1,\sigma
}+a_{I+1,1,\sigma }^{+}a_{I,2,\sigma }\right] \\
&  \\
& -t\sum\limits_{I\neq J,\sigma }\sum\limits_{\alpha =1}^2a_{I,\alpha
,\sigma }^{+}a_{J,\alpha ,\sigma }.
\end{array}
\end{equation}
Going back to the original lattice (cf (4),(5) and (1)) it is easy to see
that the exact Hubbard Hamiltonian in the limit $U\rightarrow \infty $ takes
on the form

\begin{equation}
\overline{H}=-t\sum\limits_{i,j,\sigma }a_{i,\sigma }^{+}a_{j,\sigma }
\end{equation}
where $i$ and $j$ (as before) number the lattice points and $a_{i,\sigma }$ $%
(a_{i,\sigma }^{+})$ are the Hubbard operators, defined by (9).

\section{Conclusions}

Using a new, nonperturbative approach, basing on the equivalent form of the
Hubbard Hamiltonian, represented by the collection of interacting dimers (4)
where each dimer problem has been exactly solved, we have expressed the
annihilation (creation) operators (B.1)-(B.4) as linear combinations of
transition operators between different dimer states (dimer representation).
This method made it possible to exclude from the considerations the
unoccupied dimer energy levels in the large $U$ limit by means of the
projection technique resulting in the final Hamiltonian for the dimer itself
(14), as well as, for the crystal as a whole (see (23), (24) and (C.6)). It
is important to stress that the elimination of the unoccupied dimer energy
levels was the only assumption to derive the Hubbard Hamiltonian (24) in the
large $U$ limit. Therefore we can be sure that expanding (24) with respect
to $x=\frac tU$ (Taylor expansion) we obtain absolutely all terms
proportional to $x^n$ $(n=1,2,...).$ In other words all the coefficients
proportional to $x^n$ are easy to control what is not always the case when
using another methods. The final form of the obtained Hamiltonian in the
large $U$ limit (see (23), (24) and (C.6)) visualizes high degree of
complexity of the model, deeply hidden in its original version, forming a
mixture of many multiple ferromagnetic, antiferromagnetic and more complex
interactions competing one with another. This fact seems to be decisive for
our final conclusion. Because we are still dealing with the approximate
solutions of the model (the exact solution does not exist till now) it may
happen that we underestimate in this way some important interactions and
overestimate the others. It is the reason why the resulting thermodynamic
properties of the Hubbard model, obtained in an approximate way, so strongly
depend on the quality of applied approximations.

We note in passing that the exact dimer solution may serve as a novel alloy
analogy for the Hubbard model, which could be treated by coherent potential
approximation (CPA). It is well-known that the standard alloy analogy, based
on the atomic limit, does not allow for ferromagnetism in the Hubbard model.
This may change by application of the dimer solution which already accounts
for a restricted hopping of the band electrons. A corresponding study is in
preparation.

Another example of a dimer approach to Hubbard-like models is the bond
operator theory as an extension of the slave bosonic and fermionic operators
(see Refs [40], [41] and papers cited therein).

\pagebreak

{\LARGE Appendix A}

\bigskip

The exact solution of the dimer eigenvalue problem (5) reads:

\[
\begin{array}{ll}
E_0=0; & |E_0\rangle =|0\rangle ,
\end{array}
\]

\[
\begin{array}{ll}
E_{11}=-t; & |E_{11}\rangle =\frac{1}{\sqrt{2}}(|11\rangle +|13\rangle ), \\
E_{12}=t; & |E_{12}\rangle =\frac{1}{\sqrt{2}}(|11\rangle -|13\rangle ), \\
E_{13}=-t; & |E_{13}\rangle =\frac{1}{\sqrt{2}}(|12\rangle +|14\rangle ), \\
E_{14}=t; & |E_{14}\rangle =\frac{1}{\sqrt{2}}(|12\rangle -|14\rangle ),
\end{array}
\]

\begin{equation}
\begin{array}{ll}
E_{21}=0; & |E_{21}\rangle =\frac 1{\sqrt{2}}(|23\rangle +|24\rangle ), \\
E_{22}=U; & |E_{22}\rangle =\frac 1{\sqrt{2}}(|21\rangle -|26\rangle ), \\
E_{23}=C+\frac U2; & |E_{23}\rangle =a_1(|21\rangle +|26\rangle
)-a_2(|23\rangle -|24\rangle ), \\
E_{24}=-C+\frac U2; & |E_{24}\rangle =a_2(|21\rangle +|26\rangle
)+a_1(|23\rangle -|24\rangle ), \\
E_{25}=0; & |E_{25}\rangle =|22\rangle , \\
E_{26}=0; & |E_{26}\rangle =|25\rangle ,
\end{array}
\tag{A.1}
\end{equation}

\[
\begin{array}{ll}
E_{31}=t+U; & |E_{31}\rangle =\frac 1{\sqrt{2}}(|31\rangle +|33\rangle ), \\
E_{32}=-t+U; & |E_{32}\rangle =\frac 1{\sqrt{2}}(|31\rangle -|33\rangle ),
\\
E_{33}=t+U; & |E_{33}\rangle =\frac 1{\sqrt{2}}(|32\rangle +|34\rangle ), \\
E_{34}=-t+U; & |E_{34}\rangle =\frac 1{\sqrt{2}}(|32\rangle -|34\rangle ),
\end{array}
\]

\[
\begin{array}{ll}
E_4=2U; & |E_4\rangle =|4\rangle
\end{array}
\]
where

\begin{equation}
C=\sqrt{{\left( \frac U2\right) }^2+4t^2},  \tag{A.2}
\end{equation}

\begin{equation}
a_1=\frac 12\sqrt{1+\frac U{2C}},  \tag{A.3}
\end{equation}

\begin{equation}
a_2=\frac 12\sqrt{1-\frac U{2C}}.  \tag{A.4}
\end{equation}

\pagebreak

{\LARGE Appendix B}

\bigskip

The exact dimer representation of the construction operators is given by the
following expressions:

\begin{equation}
\begin{array}[t]{ll}
c_{1,\uparrow }= & \frac 1{\sqrt{2}}\left( P_{0,11}+P_{0,12}\right) +\frac 1{%
\sqrt{2}}P_{11,25}-\frac 1{\sqrt{2}}P_{12,25} \\
& +\frac 12\left( P_{13,21}+P_{13,22}\right) +\frac 1{\sqrt{2}}\left(
bP_{13,23}+aP_{13,24}\right) \\
& -\frac 12\left( P_{14,21}-P_{14,22}\right) +\frac 1{\sqrt{2}}\left(
aP_{14,23}-bP_{14,24}\right) \\
& +\frac 12\left( P_{21,33}-P_{21,34}\right) -\frac 12\left(
P_{22,33}+P_{22,34}\right) \\
& +\frac 1{\sqrt{2}}\left( aP_{23,33}+bP_{23,34}\right) -\frac 1{\sqrt{2}%
}\left( bP_{24,33}-aP_{24,34}\right) \\
& +\frac 1{\sqrt{2}}\left( P_{26,31}-P_{26,32}\right) +\frac 1{\sqrt{2}%
}P_{31,4}+\frac 1{\sqrt{2}}P_{32,4},
\end{array}
\tag{B.1}
\end{equation}

\begin{equation}
\begin{array}[t]{ll}
c_{1,\downarrow }= & \frac 1{\sqrt{2}}\left( P_{0,13}+P_{0,14}\right) +\frac
1{\sqrt{2}}P_{13,26}-\frac 1{\sqrt{2}}P_{14,26} \\
& +\frac 12\left( P_{11,21}-P_{11,22}\right) -\frac 1{\sqrt{2}}\left(
bP_{11,23}+aP_{11,24}\right) \\
& -\frac 12\left( P_{12,21}+P_{12,22}\right) -\frac 1{\sqrt{2}}\left(
aP_{12,23}-bP_{12,24}\right) \\
& -\frac 12\left( P_{21,31}-P_{21,32}\right) -\frac 12\left(
P_{22,31}+P_{22,32}\right) \\
& +\frac 1{\sqrt{2}}\left( aP_{23,31}+bP_{23,32}\right) -\frac 1{\sqrt{2}%
}\left( bP_{24,31}-aP_{24,32}\right) \\
& -\frac 1{\sqrt{2}}\left( P_{25,33}-P_{25,34}\right) -\frac 1{\sqrt{2}%
}P_{33,4}-\frac 1{\sqrt{2}}P_{34,4},
\end{array}
\tag{B.2}
\end{equation}

\begin{equation}
\begin{array}[t]{ll}
c_{2,\uparrow }= & \frac 1{\sqrt{2}}\left( P_{0,11}-P_{0,12}\right) -\frac 1{%
\sqrt{2}}P_{11,25}-\frac 1{\sqrt{2}}P_{12,25} \\
& -\frac 12\left( P_{13,21}+P_{13,22}\right) +\frac 1{\sqrt{2}}\left(
bP_{13,23}+aP_{13,24}\right) \\
& -\frac 12\left( P_{14,21}-P_{14,22}\right) -\frac 1{\sqrt{2}}\left(
aP_{14,23}-bP_{14,24}\right) \\
& -\frac 12\left( P_{21,33}+P_{21,34}\right) +\frac 12\left(
P_{22,33}-P_{22,34}\right) \\
& +\frac 1{\sqrt{2}}\left( aP_{23,33}-bP_{23,34}\right) -\frac 1{\sqrt{2}%
}\left( bP_{24,33}+aP_{24,34}\right) \\
& -\frac 1{\sqrt{2}}\left( P_{26,31}+P_{26,32}\right) +\frac 1{\sqrt{2}%
}P_{31,4}-\frac 1{\sqrt{2}}P_{32,4},
\end{array}
\tag{B.3}
\end{equation}

\begin{equation}
\begin{array}[t]{ll}
c_{2,\downarrow }= & \frac 1{\sqrt{2}}\left( P_{0,13}-P_{0,14}\right) -\frac
1{\sqrt{2}}P_{13,26}-\frac 1{\sqrt{2}}P_{14,26} \\
& -\frac 12\left( P_{11,21}-P_{11,22}\right) -\frac 1{\sqrt{2}}\left(
bP_{11,23}+aP_{11,24}\right) \\
& -\frac 12\left( P_{12,21}+P_{12,22}\right) +\frac 1{\sqrt{2}}\left(
aP_{12,23}-bP_{12,24}\right) \\
& +\frac 12\left( P_{21,31}+P_{21,32}\right) +\frac 12\left(
P_{22,31}-P_{22,32}\right) \\
& +\frac 1{\sqrt{2}}\left( aP_{23,31}-bP_{23,32}\right) -\frac 1{\sqrt{2}%
}\left( bP_{24,31}+aP_{24,32}\right) \\
& +\frac 1{\sqrt{2}}\left( P_{25,33}+P_{25,34}\right) -\frac 1{\sqrt{2}%
}P_{33,4}+\frac 1{\sqrt{2}}P_{34,4}
\end{array}
\tag{B.4}
\end{equation}
where

\begin{equation}
P_{\alpha ,\beta }=|E_{\alpha} \rangle \langle E_{\beta} |  \tag{B.5}
\end{equation}
and

\begin{equation}
\begin{array}{l}
a=a_1+a_2 \\
b=a_1-a_2.
\end{array}
\tag{B.6}
\end{equation}
To obtain the annihilation operators of the $I-$th dimer, the index $I$
should be added $(c_{1(2),\sigma }\longrightarrow c_{I,1(2),\sigma }$, $%
P_{\alpha ,\beta }\rightarrow P_{\alpha ,\beta }^{(I)})$ in (B.1)-(B.4).

\pagebreak

{\LARGE Appendix C}

\bigskip

To present the Hamiltonian $\overline{\overline{H}}$ in a compact form we
first introduce the following operators $\left( \mu ,\nu =1,2\right) $ :

\[
\begin{array}{ll}
s_{I,\mu ;J,\nu }^{+}= & a_{I,\mu ,\uparrow }^{+}a_{J,\nu ,\downarrow }, \\
&  \\
s_{I,\mu ;J,\nu }^{-}= & a_{J,\nu ,\downarrow }^{+}a_{I,\mu ,\uparrow }, \\
&  \\
n_{I,\mu ;J,\nu ;\sigma }= & a_{I,\mu ,\sigma }^{+}a_{J,\nu ,\sigma }, \\
&  \\
n_{I,\mu ;J,\nu }= & \sum\limits_{\sigma} n_{I,\mu ;J,\nu ;\sigma }, \\
&  \\
s_{I,\mu ;J,\nu }^z= & \frac 12\left( n_{I,\mu ;J,\nu ;\uparrow }-n_{I,\mu
;J,\nu ;\downarrow }\right) ; \\
&
\end{array}
\]

\begin{equation}
\begin{array}{ll}
\underline{s}_{I,\mu ;J,\nu }^{+}= & \underline{\underline{a}}_{I,\mu
,\uparrow }^{+}a_{J,\nu ,\downarrow }, \\
&  \\
\underline{s}_{I,\mu ;J,\nu }^{-}= & a_{J,\nu ,\downarrow }^{+}\underline{%
\underline{a}}_{I,\mu ,\uparrow }, \\
&  \\
\underline{n}_{I,\mu ;J,\nu ;\sigma }= & \underline{\underline{a}}_{I,\mu
,\sigma }^{+}a_{J,\nu ,\sigma }, \\
&  \\
\underline{n}_{I,\mu ;J,\nu }= & \sum\limits_{\sigma} \underline{n}_{I,\mu
;J,\nu ;\sigma }, \\
&  \\
\underline{s}_{I,\mu ;J,\nu }^z= & \frac 12\left( \underline{n}_{I,\mu
;J,\nu ;\uparrow }-\underline{n}_{I,\mu ;J,\nu ;\downarrow }\right) ; \\
&
\end{array}
\tag{C.1}
\end{equation}

\[
\begin{array}{ll}
\underline{\underline{s}}_{I,\mu ;J,\nu }^{+}= & a_{I,\mu ,\uparrow }^{+}%
\underline{\underline{a}}_{J,\nu ,\downarrow }, \\
&  \\
\underline{\underline{s}}_{I,\mu ;J,\nu }^{-}= & \underline{\underline{a}}%
_{J,\nu ,\downarrow }^{+}a_{I,\mu ,\uparrow }, \\
&  \\
\underline{\underline{n}}_{I,\mu ;J,\nu ;\sigma }= & a_{I,\mu ,\sigma }^{+}%
\underline{\underline{a}}_{J,\nu ,\sigma }, \\
&  \\
\underline{\underline{n}}_{I,\mu ;J,\nu }= & \sum\limits_{\sigma}
\underline{\underline{n}}_{I,\mu ;J,\nu ;\sigma }, \\
&  \\
\underline{\underline{s}}_{I,\mu ;J,\nu }^z= & \frac 12\left( \underline{%
\underline{n}}_{I,\mu ;J,\nu ;\uparrow }-\underline{\underline{n}}_{I,\mu
;J,\nu ;\downarrow }\right) .
\end{array}
\]

The operators $s_{I,\mu ;J,\nu }^{z,\pm }$, $\underline{s}_{I,\mu ;J,\nu
}^{z,\pm }$, $\underline{\underline{s}}_{I,\mu ;J,\nu }^{z,\pm }$, are not
strictly the spin operators, they, however, show some similarities to the
true spin operators (as e.g. (11)-(13)) and therefore they can be called
''hopping spin operators''. Moreover, the following abreviations have to be
used:

\begin{equation}
Q\left( I,\mu ;J,\nu \right) =\overrightarrow{S}_{I,\mu }\cdot \left(
\overrightarrow{\underline{\underline{s}}}_{I,\overline{\mu };J,\nu }+%
\stackrel{\rightarrow }{\underline{s}}_{J,\nu ;I,\overline{\mu }}\right) ,
\tag{C.2}
\end{equation}

\begin{equation}
R\left( I,\mu ;J,\nu \right) =\stackrel{\rightarrow }{S}_{I,\mu }\cdot
\stackrel{\rightarrow }{S}_{J,\nu }n_{I,\overline{\mu };J,\overline{\nu }},
\tag{C.3}
\end{equation}

\begin{equation}
R^{z,\pm }\left( I,\mu ;J,\nu \right) =\left( S_{I,\mu }^zS_{J,\nu }^{\pm
}-S_{I,\mu }^{\pm }S_{J,\nu }^z\right) s_{I,\overline{\mu };J,\overline{\nu }%
}^{\mp },  \tag{C.4}
\end{equation}

\begin{equation}
R^{-,+}\left( I,\mu ;J,\nu \right) =\left( S_{I,\mu }^{-}S_{J,\nu
}^{+}-S_{I,\mu }^{+}S_{J,\nu }^{-}\right) s_{I,\overline{\mu };J,\overline{%
\nu }}^z.  \tag{C.5}
\end{equation}

The Hamiltonian $\overline{\overline{H}}$ (cf (24)) with the use of (C.1) -
(C.5) takes on the form

\[
\begin{array}{ll}
\overline{\overline{H}}= & \sum\limits_I\overline{H}_I^d-t\sum\limits_{I,%
\sigma }\left[ \underline{\underline{a}}_{I,2,\sigma }^{+}\underline{%
\underline{a}}_{I+1,1,\sigma }+\underline{\underline{a}}_{I+1,1,\sigma }^{+}%
\underline{\underline{a}}_{I+1,2,\sigma }\right] \\
&  \\
& -2t\beta \sum\limits_I\left[ Q\left( I,1;I+1,1\right) +Q\left(
I+1,2;I,2\right) \right] \\
&  \\
& +2t\delta \sum\limits_I\left[ Q\left( I,2;I+1,1\right) +Q\left(
I+1,1;I,2\right) \right]
\end{array}
\]

\[
\begin{array}{ll}
-t\beta ^2\sum\limits_I & [R\left( I,1;I+1,2\right) +R^{z,-}\left(
I,1;I+1,2\right) \\
&  \\
& +R^{z,+}\left( I,1;I+1,2\right) +R^{-,+}\left( I,1;I+1,2\right) \\
&  \\
& +R\left( I+1,2;I,1\right) +R^{z,-}\left( I+1,2;I,1\right) \\
&  \\
& +R^{z,+}\left( I+1,2;I,1\right) +R^{-,+}\left( I+1,2;I,1\right) ]
\end{array}
\]

\[
\begin{array}{ll}
-t\delta ^2\sum\limits_I & [R\left( I,2;I+1,1\right) +R^{z,-}\left(
I,2;I+1,1\right) \\
&  \\
& +R^{z,+}\left( I,2;I+1,1\right) +R^{-,+}\left( I,2;I+1,1\right) \\
&  \\
& +R\left( I+1,1;I,2\right) +R^{z,-}\left( I+1,1;I,2\right) \\
&  \\
& +R^{z,+}\left( I+1,1;I,2\right) +R^{-,+}\left( I+1,1;I,2\right) ]
\end{array}
\]

\begin{equation}
\begin{array}{ll}
+t\beta \delta \sum\limits_I\sum\limits_{\mu =1}^2 & [R\left( I,\mu
;I+1,\mu \right) +R^{z,-}\left( I,\mu ;I+1,\mu \right) \\
&  \\
& +R^{z,+}\left( I,\mu ;I+1,\mu \right) +R^{-,+}\left( I,\mu ;I+1,\mu \right)
\\
&  \\
& +R\left( I+1,\mu ;I,\mu \right) +R^{z,-}\left( I+1,\mu ;I,\mu \right) \\
&  \\
& +R^{z,+}\left( I+1,\mu ;I,\mu \right) +R^{-,+}\left( I+1,\mu ;I,\mu
\right) ]
\end{array}
\tag{C.6}
\end{equation}

\[
\begin{array}{ll}
& -t\sum\limits_{I\neq J,\sigma }\sum\limits_{\mu =1}^2\underline{%
\underline{a}}_{I,\mu ,\sigma }^{+}\underline{\underline{a}}_{J,\mu ,\sigma }
\\
&  \\
& -2t\beta \sum\limits_{I\neq J}\sum\limits_{\mu =1}^2Q\left( I,\mu ;J,%
\overline{\mu }\right) +2t\delta \sum\limits_{I\neq J}\sum\limits_{\mu
=1}^2Q\left( I,\mu ;J,\mu \right)
\end{array}
\]

\[
\begin{array}{ll}
& -t\left( \beta ^2+\delta ^2\right) \sum\limits_{I\neq J}\sum\limits_{\mu
=1}^2[R\left( I,\mu ;J,\mu \right) +R^{z,-}\left( I,\mu ;J,\mu \right) \\
&  \\
& +R^{z,+}\left( I,\mu ;J,\mu \right) +R^{-,+}\left( I,\mu ;J,\mu \right) ]
\\
&  \\
& +2t\beta \delta \sum\limits_{I\neq J}\sum\limits_{\mu =1}^2[R\left(
I,\mu ;J,\overline{\mu }\right) +R^{z,-}\left( I,\mu ;J,\overline{\mu }%
\right) \\
&  \\
& +R^{z,+}\left( I,\mu ;J,\overline{\mu }\right) +R^{-,+}\left( I,\mu ;J,%
\overline{\mu }\right) ]
\end{array}
\]
where $\overline{H}_I^d$ is given by (14). It should also be noted that the
operators $\underline{\underline{a}}_{I,\alpha ,\sigma }$ $(\underline{%
\underline{a}}_{I,\alpha ,\sigma }^{+}),$ entering in (C.1) and (C.2), are
in fact, dependent on $\beta $ and $\delta $ (see (29)). The decomposition
of the Hamiltonian $\overline{\overline{H}}$ (C.6) according to the terms
proportional to $\beta ,$ $\delta $, $\beta ^2$,$\delta ^2$ and $\beta
\delta $ has thus only a formal character in order to keep the presentation
of the Hamiltonian $\overline{\overline{H}}$ in a compact form.

The formula (C.6) is the complete expression.

\end{document}